\documentclass{sig-alternate-10pt}
\usepackage{etoolbox}
\makeatletter
\patchcmd{\@sect}{\vskip -12pt}{\vskip 0pt}{}{}
\makeatother
\usepackage[margin=1in]{geometry}
\usepackage[english]{babel}

\usepackage[colorinlistoftodos]{todonotes}
\usepackage{listings}

\usepackage{pifont}
\usepackage[T1]{fontenc}          
\usepackage[utf8]{inputenc}
\usepackage{balance}
\usepackage{subcaption}
\usepackage{booktabs}
\usepackage[inline]{enumitem}

\usepackage{amsmath,amssymb,mathtools}
\usepackage{url}
\usepackage{tabularx, makecell}
\usepackage{pdflscape}

\usepackage{graphicx}

\usepackage{tabularx}
\usepackage{graphicx}
\usepackage{rotating}
\usepackage{adjustbox}
\usepackage{multirow}
\usepackage{tikz}
\usepackage[table,xcdraw]{xcolor}    
\definecolor{prioUp}{RGB}{220,38,38}      % red
\definecolor{prioDown}{RGB}{37,99,235}    % blue
\definecolor{prioMed}{RGB}{234,179,8}     % amber

\usetikzlibrary{arrows.meta,positioning,fit,tikzmark,calc}
\newcommand{\Yes}{\checkmark}   % ✔
\newcommand{\No}{\(\times\)}    % ✗
\newcommand{\Partial}{{\color{black}\ensuremath{\circ}}} 
\newcommand{\ttfunc}[1]{\texttt{#1}} 
\renewcommand{\arraystretch}{1.7}
\usepackage[most]{tcolorbox}
\tcbset{colframe=red,colback=white,boxrule=0.9pt,arc=2pt,boxsep=1.5pt}
\newtcbox{\cellbox}{enhanced,sharp corners,boxrule=0.9pt,arc=2pt,
                    colframe=red,colback=white,boxsep=1.5pt,
                    width=\linewidth}
\usepackage[hidelinks]{hyperref}
\usepackage{rotating}
\usepackage{cleveref}
\crefname{lstlisting}{listing}{listings}
\Crefname{lstlisting}{Listing}{Listings}

\begin{document}
\pagestyle{empty}
% === listing-style.tex =================================================
% Assumes xcolor + listings are already loaded in the main preamble.
% (acmart loads hyperref; do NOT load geometry/hyperref here)

% ---- palette (yours) ----
\definecolor{keyword}{HTML}{2771a3}
\definecolor{keyword-new}{HTML}{27a330}
\definecolor{pattern}{HTML}{b53c2f}
\definecolor{string}{HTML}{be681c}
\definecolor{relation}{HTML}{7e4894}
\definecolor{variable}{HTML}{107762}
\definecolor{comment}{HTML}{8d9094}

% ---- default style for ALL code blocks (no language set here) ----
\lstdefinestyle{codeblock}{
  basicstyle=\small\ttfamily,
  commentstyle=\color{comment}\ttfamily,
  keywordstyle=\color{keyword}\bfseries\ttfamily,
  stringstyle=\color{string}\ttfamily,
  numbers=left,
  numberstyle=\tiny\ttfamily,
  numbersep=6pt,
  showstringspaces=false,
  aboveskip=3pt,
  belowskip=3pt,
  columns=fullflexible,       % no big gaps
  keepspaces=true,
  breaklines=true,
  tabsize=2,
  frame=single,
  captionpos=b,
}
\lstset{style=codeblock, upquote=true}

% ---- Cypher language ----
\lstdefinelanguage{cypher}{
  keywords=[1]{MATCH,OPTIONAL,WHERE,NOT,AND,OR,XOR,RETURN,DISTINCT,ORDER,BY,
    ASC,ASCENDING,DESC,DESCENDING,UNWIND,AS,UNION,WITH,ALL,CREATE,DELETE,DETACH,
    REMOVE,SET,MERGE,SKIP,LIMIT,IN,CASE,WHEN,THEN,ELSE,END,INDEX,DROP,UNIQUE,
    CONSTRAINT,EXPLAIN,PROFILE,START,GRAPH,CONSTRUCT,STREAM,FROM,LOAD},
  keywordstyle=[1]\color{keyword}\bfseries\ttfamily,
  keywords=[2]{REGISTER,QUERY,RSTREAM,DSTREAM,ISTREAM,WINDOW,RANGE,SLICE,OUTPUT,
    EVERY,Hours,Minutes,Seconds,STARTING,Earliest,Latest,EMIT,ON,EXIT,ENTERING,
    INTO,SNAPSHOT,WITHIN,AT,REPORTING},
  keywordstyle=[2]\color{keyword-new}\bfseries\ttfamily,
  % only for cypher snippets:
  literate=*
    {...}{{\cdot\!\cdot\!\cdot}}{1}
    {theta}{{$\theta$}}{1}
}

% ---- JSON language (minimal, robust) ----
\lstdefinelanguage{json}{
  morestring=[b]",                      % strings
  stringstyle=\color{string},
  keywords={true,false,null},           % literals
  keywordstyle=\color{keyword}\bfseries,
  % punctuation colouring
  literate=
    *{:}{{\textcolor{relation}{:}}}{1}
     {,}{{\textcolor{relation}{,}}}{1}
     {\{}{{\textcolor{relation}{\{}}}{1}
     {\}}{{\textcolor{relation}{\}}}}{1}
     {[}{{\textcolor{relation}{[}}}{1}
     {]}{{\textcolor{relation}{]}}}{1},
}

% ---- Flux language + example style (optional) ----
\lstdefinelanguage{Flux}{
  morekeywords={
    from,range,filter,group,last,first,mean,sum,count,keep,drop,rename,
    pivot,join,union,distinct,sort,limit,map,reduce,aggregateWindow,
    toInt,toFloat,toBool,time,now,option,with,tables,columns,method,on
  },
  sensitive=true,
  morecomment=[l]{//},
  morestring=[b]"
}
\lstdefinestyle{fluxstyle}{
  language=Flux,
  basicstyle=\ttfamily\small,
  keywordstyle=\bfseries\color{blue!70!black},
  commentstyle=\itshape\color{gray!70!black},
  stringstyle=\color{green!40!black},
  numbers=left,
  numberstyle=\tiny\color{gray},
  numbersep=8pt,
  frame=single, framerule=0.3pt,
  columns=fullflexible, breaklines=true, tabsize=2, captionpos=b
}

% ---- MongoDB shell / aggregation language ----
\lstdefinelanguage{mongodb}{
  sensitive=true,
  alsoletter={\$},                      % let $ be part of identifiers
  morecomment=[l]{//},
  morecomment=[s]{/*}{*/},
  morestring=[b]", morestring=[b]',
  % shell/CRUD keywords
  keywordstyle=\color{keyword}\bfseries\ttfamily,
  keywords={db,use,show,quit,exit,find,findOne,insert,insertOne,insertMany,
    update,updateOne,updateMany,replaceOne,deleteOne,deleteMany,aggregate,
    count,countDocuments,distinct,createIndex,createIndexes,drop,dropIndex,
    dropIndexes,dropDatabase,explain,bulkWrite,watch,stats},
  % pipeline stages
  keywordstyle=[2]\color{keyword-new}\bfseries\ttfamily,
  keywords=[2]{\$match,\$group,\$project,\$sort,\$limit,\$skip,\$lookup,\$unwind,
    \$addFields,\$set,\$unset,\$replaceRoot,\$replaceWith,\$count,\$out,\$merge,
    \$unionWith,\$facet,\$bucket,\$bucketAuto,\$sortByCount,\$graphLookup,
    \$redact,\$sample,\$geoNear},
  % operators
  keywordstyle=[3]\color{pattern}\bfseries\ttfamily,
  keywords=[3]{\$and,\$or,\$not,\$nor,\$in,\$nin,\$eq,\$ne,\$gt,\$gte,\$lt,\$lte,
    \$exists,\$type,\$regex,\$size,\$all,\$elemMatch,\$push,\$pull,\$set,\$unset,
    \$inc,\$addToSet,\$each,\$slice,\$position,\$min,\$max,\$sum,\$avg,\$first,\$last},
  % punctuation (JSON-ish)
  literate=
    *{:}{{\textcolor{relation}{:}}}{1}
     {,}{{\textcolor{relation}{,}}}{1}
     {\{}{{\textcolor{relation}{\{}}}{1}
     {\}}{{\textcolor{relation}{\}}}}{1}
     {[}{{\textcolor{relation}{[}}}{1}
     {]}{{\textcolor{relation}{]}}}{1},
}
% ---- SQL language ----------------------------------------------------
\lstdefinelanguage{sql}{
  sensitive       = false,                 % SQL keywords are case-insensitive
  alsoother       = {\$},                   % let $ appear in identifiers
  % --- core keywords (blue) ---
  morekeywords=[1]{
    select,insert,update,delete,merge,into,values,set,from,where,group,by,having,
    order,limit,offset,fetch,distinct,all,union,intersect,except,join,inner,left,
    right,full,cross,on,using,natural,as,create,drop,alter,truncate,table,view,
    index,primary,foreign,key,constraint,unique,not,null,default,check,if,exists,
    case,when,then,else,end,between,in,like,is,and,or,xor,asc,desc,
    grant,revoke,commit,rollback,savepoint,begin,transaction
  },
  keywordstyle=[1]{\color{keyword}\bfseries\ttfamily},
  % --- data types & pseudo types (green) ---
  morekeywords=[2]{
    int,integer,smallint,bigint,numeric,decimal,float,double,real,
    serial,boolean,bit,char,varchar,text,date,time,timestamp,interval,
    json,jsonb,uuid,xml,bytea,blob,clob
  },
  keywordstyle=[2]{\color{keyword-new}\bfseries\ttfamily},
  % --- functions & operators (red) ---
  morekeywords=[3]{
    count,sum,avg,min,max,coalesce,nullif,upper,lower,substring,trim,length,
    now,current_date,current_time,current_timestamp,row_number,rank,dense_rank,
    over,partition,lag,lead,first_value,last_value
  },
  keywordstyle=[3]{\color{pattern}\bfseries\ttfamily},
  % comments & strings
  morecomment    = [l]{--},             % -- line comment
  morecomment    = [s]{/*}{*/},         % /* block comment */
  commentstyle   = \color{comment}\ttfamily,
  morestring     = [b]',                % 'single-quoted strings
  stringstyle    = \color{string},
  % misc formatting
  columns        = fullflexible,
  keepspaces     = true,
  breaklines     = true,
  literate=
    *{:}{{\textcolor{relation}{:}}}{1}
     {,}{{\textcolor{relation}{,}}}{1}
     {\{}{{\textcolor{relation}{\{}}}{1}
     {\}}{{\textcolor{relation}{\}}}}{1}
     {[}{{\textcolor{relation}{[}}}{1}
     {]}{{\textcolor{relation}{]}}}{1}
}
% ----------------------------------------------------------------------

\title{Combining Time-Series and Graph Data: A Survey of Existing Systems and Approaches \thanks{This paper is submitted to SIGMOD Record.}}
\numberofauthors{3}
\author{
\alignauthor Mouna Ammar\\
  \affaddr{University of Leipzig, ScaDS.AI}\\
  \affaddr{Leipzig, Germany}\\
  \email{ammar@informatik.uni-leipzig.de}
\and
\alignauthor Marvin Hofer\\
  \affaddr{University of Leipzig, ScaDS.AI}\\
  \affaddr{Leipzig, Germany}\\
  \email{hofer@informatik.uni-leipzig.de}
\and
\alignauthor Erhard Rahm\\
  \affaddr{University of Leipzig, ScaDS.AI}\\
  \affaddr{Leipzig, Germany}\\
  \email{rahm@uni-leipzig.de}
}

\maketitle

\pagestyle{empty}
\begin{abstract}
%In recent years, graphs and time series data have been widely used in research and industry. However, few data models support joint modeling of these two types of data.
%In this work, 
We provide a comprehensive overview of current approaches and systems for combining graphs and time series data. We categorize existing systems into four architectural categories and analyze how these systems meet different requirements and exhibit distinct implementation characteristics to support both data types in a unified manner. 
%arising from the need for a hybrid storage solution. We conclude that no survey system currently achieves the envisioned unification of both data models. 
Our overview aims to help readers understand and evaluate current options and trade-offs, such as the degree of cross-model integration, maturity, and openness.
%Deeper integration designs unify types, hybrid operators, and cross-model consistency, while shallower integration retains more advanced PG/TS features and ecosystems but weakens hybrid capabilities.
\end{abstract}
\section{Introduction}
\label{sec:introduction}

Graph and time-series data are needed in many areas and are typically managed in separate data management systems. 
Graph data is widely used, e.g., in social networks, micromobility, and finance ~\cite{patel2017graph,vicknair2010comparison, overcoming2015, pokorny2015graph}, to model complex relationships and apply analysis ranging from pattern matching and traversal to graph machine learning. Graph data is either managed in native systems such as Neo4j or JanusGraph, which are based on a property graph model, or in more general data stores such as key-value stores or relational database management systems (DBMS)~\cite{besta2023demystifying}.
%document  LPG graph store (e.g, Neo4j), document (e.g., ArangoDB), column store (e.g., Kuzu), RDBMS, key-value or RDF store~\cite{besta2023demystifying}%There are multiple graph databases, both with native implementations, such as Neo4j, and non-native ones, such as Spanner Graph. 
%Temporal graph data systems manage the evolution of graph data and support time-related analysis capabilities such as change analysis ~\cite{michail2016introduction, debrouvier2021model}.    
Time series are another essential data structure for recording measurements or events over time (e.g., stock prices, sensor readings) and for supporting time-related analytics such as prediction and anomaly detection~\cite{rani2014review,schmidl2022anomaly}. Again, this kind of data can be managed either in native stores (e.g., time-series DBMS such as TimescaleDB~\cite{timescaldb} and InfluxDB~\cite{influxdb}) or in more general data stores. 

Applications increasingly demand the combined analysis of both graph data and time series ~\cite{shao2022pre,liu2023largest,mousavi2019stanford},  notably in machine learning.  
For example, the connections (e.g., road segments) of a traffic network can be extended with time-series data to capture vehicle counts or average speeds, enabling identification of congestion and its influence on neighboring areas.  Similarly, the analysis of sensor time series in an Internet of Things (IoT) environment, e.g., for anomaly detection, can be improved by accounting for sensor topology and potential dependencies among sensors.  

Storing graph data and time-series data in separate data stores is a minimal approach for such applications, as they would have to manage different systems and query languages. Furthermore, they would have to implement all   
queries and analysis tasks that require both data types.  
Hence, we see a strong need for closer integration of graph data and time series within a single system that provides not only specific support for both data types but also new (hybrid) operations on them. A vision for such a powerful hybrid approach has been presented in ~\cite{ammar2025towards}, but a complete implementation remains lacking.

In this work, we therefore focus on surveying existing approaches for combining graph and time-series data. We categorize solutions into several architectural integration types and evaluate the different systems against a list of requirements and system characteristics.  
We consider four 
integration architectures: (i) Single model \textbf{(SM)} systems that use a single data model (either a native graph or time series model or another data model) to manage both graph and time series data, 
(ii) Extended Single Model \textbf{(E-SM)} systems that extend a main model with dedicated extensions for graph/time series support
, (iii) \textbf{Polyglot} systems integrating multiple engines via middleware, and (iv) Multi Model \textbf{(MM)} systems that do not have a single preferred data model but support multiple equally important models in one engine. 
%We scope our system to database families capable of storing both graphs and time series in practice: graph DBMSs, time-series DBMSs, document stores, relational systems, and wide column stores. 
%and the \textbf{degree of their integration}. 
%We call a system deeply integrated when a single logical data model can (a) express temporal graphs and both explicit and implicit time‑series, (b) store them without loss of transformations, and (c) support hybrid operators whose execution plans are optimized across the two domains. 
Our survey aims to help readers understand the current options and their strengths and limitations. The shown methodology can also be applied to systems not covered in this paper.  We release the codebase used to assess all four architectures for executing cross-model queries in \cite{rep_queries}.

We make the following contributions:
\begin{enumerate}
    \item We propose a set of requirements as well as relevant system characteristics for the combined management and analysis of graph and time series data 
    %to provide the needed level of integration between the two data models, and we map them into system characteristics that would be used later on as criteria for combining systems
            \item We describe four architectural approaches to combining graphs and time series, along with their advantages and disadvantages.
        %proofs of concept;
    \item We comparatively evaluate 20 systems of the different categories w.r.t. the introduced requirements and characteristics.
\end{enumerate}

The remainder of the paper is organized as follows.
\Cref{sec:relatedwork} presents related surveys,
\Cref{sec:taxonomy} outlines  the requirements and system characteristics for the comparative evaluation %for a \textbf{deep integration} and characteristics used to measure how well each surveyed system meets the requirements  
\Cref{sec:integration-type}  describes the  integration architectures.
\Cref{sec:results} surveys representative systems and research prototypes.
We synthesize the open challenges in \Cref{sec:challenges} and conclude our work in \Cref{sec:conclusion}.

%\section{Preliminaries}
%\section{From Phenomena to Practice: Importance of Hybrid Solutions in the Micromobility Context}
%\label{sec:usecase}
%\input{motivation_use-case}

\section{Related Surveys} 
\label{sec:relatedwork}

% Existing surveys have addressed the challenges of managing time-series data on the one hand and graph data on the other; several also discuss broader polyglot or multimodel systems. Polyglot persistence approaches store and query data in separate specialized engines, while multimodel databases provide single-engine support for multiple paradigms (e.g., document, relational, graph) under one system\cite{}. However, no existing survey has comprehensively explored the full integration of graphs and time series at the storage level—that is, how to natively represent and query these two data models in a single database engine.
%Surveys on graph and time-series management have mainly evolved in isolation, while broader work on multimodel and polyglot systems emphasizes architectural trade-offs. 
Multiple surveys  have been conducted on solely time-series databases~\cite{jensen2022time,jensen2017time,bader2017survey,khelifati2023tsm, jalal2022scits, visperas2021time,yuanzhe2021ts} and on graph databases~\cite{besta2023demystifying,coimbra2025survey,anuyah2024understanding, DBLP:journals/pvldb/IosupHNHPMCCSAT16, DBLP:journals/corr/abs-2307-12510,ldbc}.
Besta et al.~\cite{DBLP:journals/csur/BestaGPFPBAH24} classify more than 50 systems by data models, storage layouts, and query execution strategies, highlighting fundamental design tensions in handling dynamic, large-scale graphs. 
None of these surveys addresses approaches to supporting both time-series and graph data, as we do in this work.
There are several models and implementations for\textit{ temporal graph database systems} extending property graphs (\cite{huang2016tgraph, debrouvier2021model, hartmann2017analyzing, michail2016introduction, moffitt2017temporal, wu2014path, xu2017time}) or the RDF  graph data model  \cite{DBLP:journals/access/ZhangWLC19, DBLP:journals/esi/ZhangLPC21}.
These approaches can manage and analyze the evolution of graph data over time. However, they do not cover time series as we aim for.  
Lu and Holubová ~\cite{lu2019multi} provide a general overview of multi-model databases and a comparison of different systems, but without considering support for time series data and without taking specific requirements into account as we do here. 

 \begin{figure*}[t] 
    \centering
    \includegraphics[width=1\linewidth]{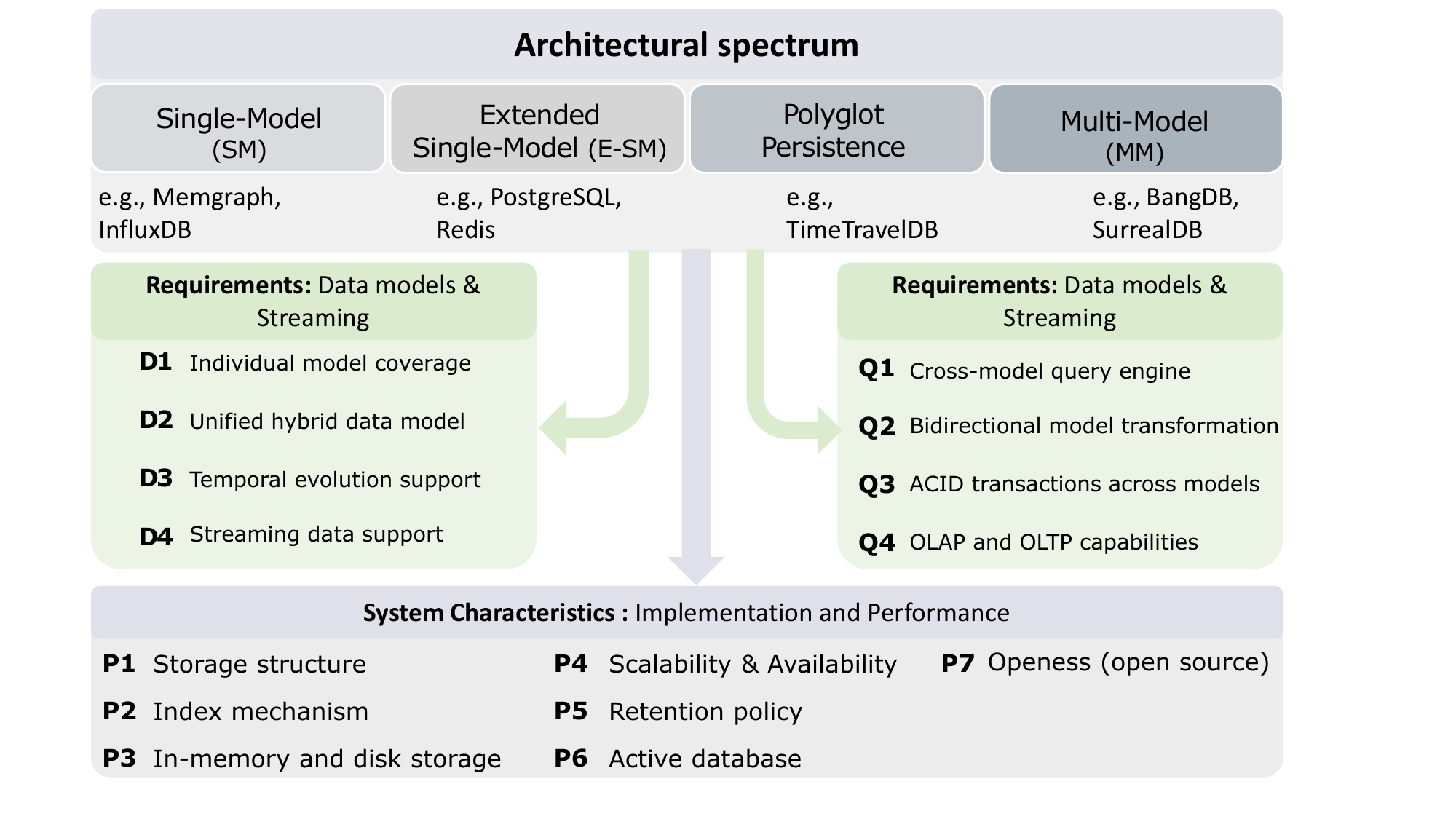}
   \caption{Overview of the comparison approach of systems for combining graph and time series data}
  \label{fig:comparison-summary}
\end{figure*}

\section{Criteria for comparison}
\label{sec:taxonomy}

%\noindent
%This section defines how we compare systems based on our methodology that comprises three main components:
% \begin{enumerate}[leftmargin=*]
%     \item 
%    (1) Classify systems by their \emph{integration type} (\Cref{sec:integration-type}),
    % \item 
%    (2) state the \emph{System requirements} (D*, Q*, SC*) that measure deep integration (\Cref{subsec:req-facets}), and
    % \item 
%    (3) list the \emph{System characteristics} (P*, M*) we use to describe how well each system meets those requirements (\Cref{subsec:char-facets}). 
% \end{enumerate}
We define a set of key requirements and desirable system characteristics for a data management system that supports both graph and time-series data.
This is essentially a wish list for a perfect solution, which current solutions will only partially cover. Still, we believe they provide a reasonable basis for comparing available systems and identifying currently missing functionality.

\subsection{System Requirements}
\label{subsec:req-facets}
%We evaluate deep integration along three requirement classes.
\noindent \textbf{Requirements Group 1:} Data model and Streaming
\begin{enumerate}[label=D\arabic*]

    \item Individual model coverage: There should be data model support for graph data, e.g., for property graphs with labeled data nodes and edges, as well as for time series data. 
    \item Unified hybrid data model: Ideally, there should be a unified data model that supports a seamless integration of both graph and time series data as first-class citizens, as well as their combined representation (e.g., time series properties of graph nodes or interlinked time series). 
    %The integration should ensure consistent representation between graphs and time series.
    \item Temporal evolution: Support is needed for both current and historical data to capture how graphs and time series evolve over time. 
    %For graphs, through storing when graph elements are valid; providing historical graph states. For timeseries, through handling continuously growing series; out-of-order and late-arriving data and managing updates~\cite{wang2025apache}  (e.g., append-only logs, versioning, bitemporal timestamps)
   \item Streaming data support: The system should support ingestion, processing, and querying of unbounded high-velocity data streams, in particular continuously arriving time series and graph updates, with low-latency processing, proper handling of out-of-order or late events, and integration with stored state~\cite{stonebraker20058, watanabe2007integrating}

    \end{enumerate}
    
     \noindent \textbf{Requirements  Group 2:} Queries and transformations
    \begin{enumerate}[label=Q\arabic*] 

 \item Cross‑model query engine: 
 %Executes queries that mix graph and time‑series operators in a single statement. 
 It should be possible to execute queries on both graph and time-series data, and to mix them in a single query. Users must be able to join,
filter, aggregate, and pattern-match across the two models without
manually orchestrating multiple query engines. In addition to hybrid queries (e.g., time-aware grouping), support should also include extraction operators (e.g., extracting time series from graph data) and hybrid updates that span both data models with consistent semantics. Ideally, this capability is
exposed through a unified query abstraction or query language~\cite{guo2024multi, koupil2023mm}

    % \item Unified query language:
    %and/or hybrid operator algebra: 
   %In the case of a hybrid data model (D2), there should be a unified query language that supports both graph and time-series data. 
%    \item Hybrid data ingestion: import a dataset that combines graph and time-series data, preserving links to ensure that fresh data is quickly available for queries (timeliness).
     \item Bidirectional model transformation  (interoperability and flexibility): Support is needed to transform data between the single data models (from graph to time series and back) as well as between integrated hybrid model and single-model (graph-only or time-series-only) formats, as well as for import and export of single‑model and hybrid data. 
    \item ACID transactions across data types: Atomic, consistent updates that span graph and time‑series entities in the same transaction.
      \item OLTP and OLAP capabilities: Support for both transactional OLTP workloads (point queries, updates, pattern matching, snapshot queries) and analytical OLAP workloads (aggregations, global graph and time series algorithms) without requiring separate ETL~\cite{attobrah2024etl} or systems. Such hybrid HTAP capabilities ~\cite{li2022htap} are beneficial because hybrid graph and time series applications inherently combine transactional operations (e.g., graph traversal) with analytical operators (e.g., time series aggregations).
    \end{enumerate}
%  \paragraph{\textbf{Criteria Group 3:} 
%  Optimized implementation}
%  The posed data model and query requirements are challenging to meet and ask for advanced implementation techniques to achieve fast processing and scalability to large datasets. While we could have defined a long list of implementation requirements we only mention three aspects but will discuss more general system characteristics below.  
%    \begin{enumerate}[label=SC\arabic*] 
%    \item Flexible storage support for hybrid data: Capability of storing graph, time-series, and their linkage either jointly in a unified store or separately in dedicated engines.
%      \item In-memory and disk storage: Storing and processing data in-memory for fast analysis and on disk for persistence and durability. 

     % \item Scalability: Ability to execute low-latency point queries or updates and big analytical jobs on the integrated dataset efficiently. 
%    \end{enumerate}          

\subsection{System Characteristics}
\label{subsec:char-facets}
The posed data model and query requirements are challenging to meet and require advanced implementation techniques to achieve fast processing and scalability with large datasets. Given that multiple system options partially meet the posed requirements, we do not specify implementation requirements. Instead, we identify relevant system characteristics related to implementation and performance that influence how well the posed requirements can be met. These aspects will also be considered in our comparative evaluation of current implementations.  
%following are the implementation dimensions we score to determine how well a system satisfies the requirements outlined in \Cref{subsec:req-facets}.
%(3) list the \emph{System characteristics} (P*, M*) we use to describe how well each system meets those requirements (\Cref{subsec:char-facets}). 
%\paragraph{\textbf{Characteristics Group 1:} Modeling and Query Processing}
%\begin{enumerate}[label=M\arabic*, leftmargin=1.6em]
%\item Core database model. E.g., graph, time-series, relational, document, key–value, wide-column.
% \item Single model support. Degree of native support for (property) graphs and time series.
% \item Temporal data support, e.g., using snapshots or validity intervals for entities and relationships 
%  \item Query interface and language: Indicates how users express graph, time series, and cross-model queries.
% \item Supported graph and time series operators.
%  \item Hybrid support \& hybrid data updates. Hybrid support refers to Q3 (bidirectional model transformations), in which both data models are treated as first-class entities and linked. The link ensures consistent representation between graphs and time series. 
% As for hybrid data updates, this indicates the ability to support updates/ingestion that span both data models while maintaining consistent semantics.
% \item Data model transformation. Built-in mechanisms to project one model to another while preserving structure and semantics.
% \item Hybrid ingestion: Single-pass ingestion that loads graph elements, time-series, and their linkages atomically.
%\end{enumerate}

\noindent \textbf{Implementation and Performance Characteristics}
\begin{enumerate}[label=P\arabic*, leftmargin=1.6em]

\item Storage structure of graph and TS data,  e.g., adjacency matrix, B$^{+}$-tree, log-structured merge (LSM) tree, columnar storage. 
       \item Graph-TS Index mechanisms:  Indexing support for graph and time series data as well as for composite indexing, e.g., to jointly capture graph entities with their time-series attributes.
       \item In-memory and disk storage: Support for in-memory storage and processing (e.g., for fast analysis) and external storage for persistence and durability. 
     \item Scalability and availability mechanisms: Support horizontal scaling through data distribution (e.g., sharding) to handle growing volumes of graph and time-series data while maintaining query performance and maintaining fault tolerance through replication.
          \item Retention policy. Built-in lifecycle rules (e.g., time-to-live, downsampling) of historical data control how long graph and time-series data are retained. 
               \item Active database support to react to changes, e.g., ECA rules (Event-Condition-Action), triggers, or standing/continuous queries~\cite {margara2014streaming}.
  \item Openness. Commercial vs. open-source license. Systems released under an OSI-approved license~\cite{opensource} (e.g., Apache 2.0, GPL, PostgreSQL license, MIT, BSD) are considered open-source systems. Other systems under licenses such as SSPL/BSL, with or without available source, are considered closed-source.
\end{enumerate}

   Figure \ref{fig:comparison-summary} summarizes the criteria and characteristics introduced in our comparison approach.

\section{Integration architectures}
\label{sec:integration-type}

\begin{figure*}  
    \centering
    \includegraphics[width=1\linewidth]{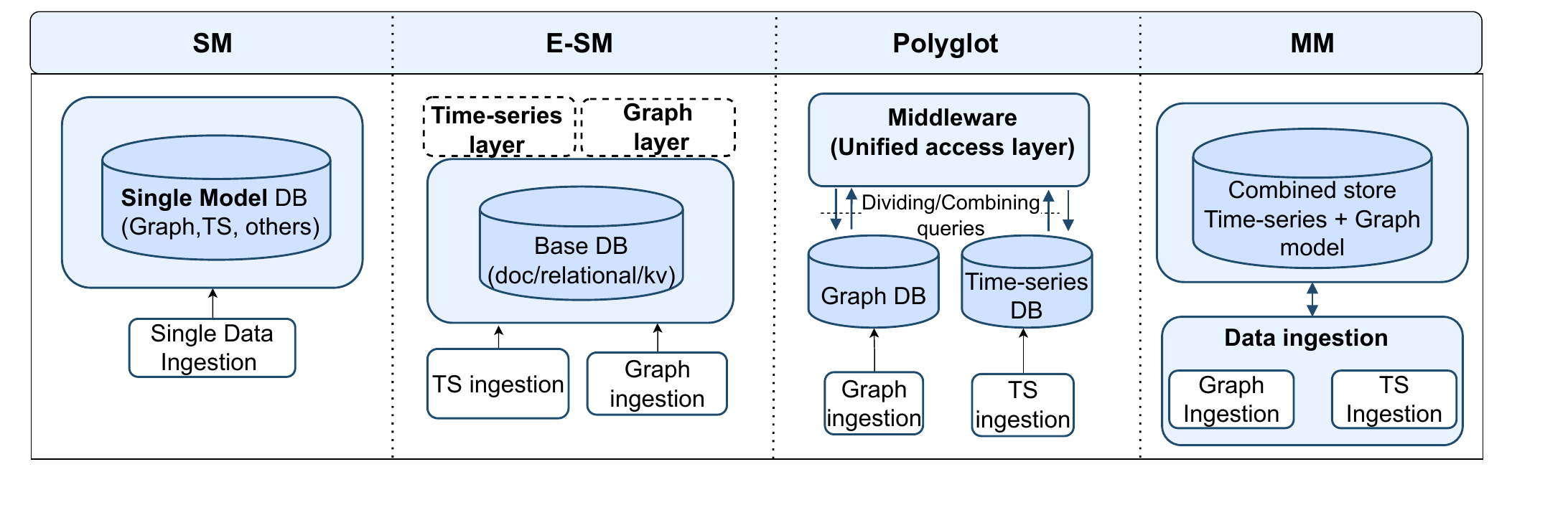}
    \caption{Sketches of integration architectures: SM, E-SM, Polyglot, and MM }
    \label{fig:architecture-sketch}
    
\end{figure*}

%We discuss the theoretical foundations and practical implications of combining graph and time series data. 
We categorize existing systems for combining graph and time series data into four architectural approaches: \emph{Single-model (SM)}, \emph{Extended Single-model (E-SM)}, \emph{Polyglot persistence system}, and  \emph{Multi-model database (MMDB)}\footnote{These architectures are similarly applicable when other kinds of data should be combined.}. 
%We also include the \emph{Unified Hybrid Model} as a visionary architecture target. \Cref {tab:architecture_comparison} summarizes the benefits and challenges of the four different architectures, providing examples. %By exploring these methodologies, we aim to identify the most effective strategies for combining graphs with time series data, which will guide the architectural design of the hybrid system.
The gross architectures of these solutions are shown in \Cref{fig:architecture-sketch}.
\Cref{tab:architecture_comparison}  summarizes some of the benefits and challenges of the four architectures and lists sample implementations. This section discusses the architectures in more general terms, while the next section will comparatively evaluate selected systems across the four integration architectures.  In our repository~\cite{rep_queries}, we provide solutions for several DBMS architectures for storing and analyzing graph and time-series data in a mobility use case (bike sharing). 

\subsection{Single-model systems (SM)}
\label{subsec:single-engine}
     This approach uses a single-model DBMS  to store data of other types. 
     %Examples of systems within this family are shown in \Cref{tab:architecture_comparison}. 
     Three variations of this approach can be distinguished:

\begin{itemize}[leftmargin=*, topsep=0pt] 
    \item \textit{Graph database systems that store time series data.} This method natively supports graph data processing, e.g., for a property graph model (e.g., Neo4j~\cite{neo4j}, JanusGraph~\cite{janusgraph}, TigerGraph~\cite{tigergraph}, Memgraph~\cite{memgraph}) or a temporal graph approach (~\cite{miao2015immortalgraph,massri2022clock} but also stores and manage time series data, e.g., within node or edge properties. 
    \item \textit{Time-series DBMS storing graph data.} This is the symmetric approach where a native time series DBMS such as InfluxDB~\cite{influxdb} aims at also storing graph data, e.g., by treating nodes and edges as time-stamped events. 
    %A time-series data model is indeed too simple to handle the complexity and constraints of a graph data model~\cite{hartmann2017analyzing}. However, it is still possible to encode graph elements through the InfluxDB data model~\cite{influxdb} by treating nodes and edges as time-stamped events. An example can be found in \Cref{app:influxdb}
    \item  \textit{Other database systems storing graph and time series.}  In this approach, a DBMS with no native support for graph or time series data, e.g., a relational or NoSQL DBMS, is used to store and process these kinds of data. In a relational DBMS (e.g., MySQL~\cite{mysql}), one can store this data in different tables and use SQL to analyze graph and time-series data, which typically have substantial limitations in performance and analysis. Object-relational DBMSs address some of these limitations through dedicated extensions for non-relational data, thereby enabling an E-SM architecture.  In document stores such as MongoDB~\cite{mongodb}), each time series point can be modeled as a document carrying a timestamp. Such documents can then be \emph{linked} (via references/foreign keys) or \emph{embedded} to form a graph-shaped structure, as detailed further in our solution available in this repository~\cite{rep_queries}. 
    %Relational databases  can model graphs using foreign-key relationships between node and edge tables or as a table of adjacency lists or edges if no attributes are needed~\cite{chen2013managing}, and time series as timestamped rows with window functions.
    Key-value stores (e.g., RockDB~\cite{rockdb}) can encode graph adjacency lists as values keyed by node identifiers and time series as sorted sets or lists.
\end{itemize}

\noindent
%\textbf{Challenges.}
Systems of type SM are typically most limited w.r.t. our posed requirements since they have dedicated support for at most one of the two data types, thereby lacking advanced query support for the other data type as well as for cross-model queries. 
%Managing and querying a combined set of graph and time series data in a single system increases complexity, particularly in ensuring that the queries are efficient and that the data are stored in a way that does not degrade performance. Depending on the database, native capabilities for handling one data type might be more developed than those for the other. For example, while Neo4j is optimized for graph operations, its handling of time-series data may require additional customization to meet specific requirements. 
Even temporal graph DBMS with query languages such as T-GQL~\cite{debrouvier2021model} and T-PGQL~\cite{rost2023evolution} 
%enrich property graph querying with temporal constructs but 
do not support time-series analysis. Similarly, non-graph DBMS have neither dedicated support for 
%MongoDB offers flexible document storage, but 
complex graph queries (e.g., k-hop pattern matching), nor for cross-model queries. 
%may not be as efficient as those in a specialized graph database. InfluxDB supports joins and range queries but lacks deep graph analytics, such as k-hop patterns, because joins become costly beyond two hops. %As data volumes grow, scaling a single system to handle both high-volume time-series data and complex graph data efficiently can be challenging.  

 \subsection{Extended single-model (E-SM)}
 \label{subsec:mme}
Systems within the E-SM architecture are based on a mature core data model, such as the relational model, but extend it (e.g., via plugins or modules) to support additional data types, including JSON, graph, and time-series data. This leads to multi-model capabilities, but the different 
models are not equally important as in the architectural type MM. Furthermore, the additional data types are often mapped to the original model's storage structures; i.e., data are ultimately stored according to the core model's paradigm, which can impose restrictions on possible queries and performance. 

The E-SM approach is explicitly supported by the object-relational extensions of the relational model, which enable user-defined data types and procedures, e.g., to support non-relational data. The SQL standard has already defined such extensions for document data (XML, JSON) and graph data \cite{deutsch2022graph} 
but not yet for time-series data (although SQL window predicates are useful for time-series analysis). The operators of the extended data types can be used in SQL queries, thereby supporting cross-model queries and data type transformations.  
Existing DBMS, such as PostgreSQL~\cite{postgres} and Oracle, provide extensions that support both graph and time-series data, thereby extending the SQL standard. 
%MariaDB also supports storing graph and time-series data, but support for cross-model queries is minimal. They require an ETL-like step to communicate between data models for query processing: exporting results from one query model and importing them into another to combine them.

The key-value store 
 Redis~\cite{redis} provides extensions for both graph and time-series data. 
% or PostgreSQL~\cite{postgres}, start with a mature core data model, such as key-value for Redis and relational for PostgreSQL, and then extend their capabilities to support additional models, like graph, JSON, or time-series, through modules, layers, or plugins~\cite{jevzekdortdb}.
%  The result is a multi-model system~\Cref {subsec:mmn}. Multi-model systems can be divided into native single-engine systems and extended single-model systems \cite{zhang2021holistic}.
%Two criteria distinguish E-SM from native multi-model systems: 
%\begin{itemize}[leftmargin=*]
%    \item The system is originally supporting one single mature model (i.e., relational, key-value, etc) and was extended using layers or plugins to support other data models;
%    \item The secondary models are mapped or translated to the original model's storage structures, i.e., data are ultimately stored according to the core model's paradigm.
%\end{itemize}
For the wide-column data store Cassandra~\cite{cassandra, planet-cassandra}, there are several options for storing both data types. Cassandra can either be used as a storage backend with both Janusgraph (for graph data)~\cite{janus4cassandra} and KairosDB~\cite{kairos} (for time series data) built on top of it, or use DataStax Graph~\cite{Datastax}, a native extension of the commercial Cassandra distribution to store graph data and time series data directly in Cassandra, which already provides optimized storage and some analysis capabilities. An example query is provided in \cite{rep_queries}. %A more complex option without using extensions would be store data models directly in Cassandra by creating a table for vertices, a table for implementing the adjacency list in a denormalized table (source node id and target node id duplicated) and time series data inside a dedicated table and use the keyword "WITH CLUSTERING ORDER BY (timestamp)" so that time series data is sorted within the partition, to enable more efficient queries.

\noindent 
%\textbf{Challenges.} 
The discussion shows that there are substantially different implementations within the E-SM category, even among relational DBMS. The E-SM approach can support critical requirements such as cross-model queries, but expressiveness and performance depend on the chosen mappings and implementations based on the core data model. 

%Despite the simplified deployment it offers by providing extensions within a single engine, it still lacks native integrations that involve providing a certain guarantee of consistency between data models. The queries often require casting before joins, and the planner cannot optimize across the combination of the \texttt{cypher()} function and SQL.

\subsection{Polyglot persistence system}
 \label{subsec:polyglot}
 
 Polyglot persistence~\cite{fowler2012nosql,wiese2015polyglot} is an architectural concept that integrates multiple data types and technologies to support applications with diverse data processing needs by leveraging the strengths of different DBMS to optimize specific data model, query, or scalability requirements~\cite{glake2022towards}. The core motivation behind polyglot persistence is that no single database can fit well all data types and workloads~\cite{khine2019review}. For our purposes, this approach could be used to integrate a graph DBMS (e.g., Neo4J) with a time-series DBMS (e.g., InfluxDB). 
 
The downside is that all cross-system features must be provided by an additional middleware layer on top of the native systems (\Cref{fig:architecture-sketch} (Polyglot)).  For query processing, cross-model queries are either to be implemented by the applications or the middleware could support specific cross-model query features or even a hybrid query language based on 
suitable cross-model schemata. In that case, the middleware would have to split cross-model queries into single-model subqueries and to combine the results for the subqueries. Obviously, this is very challenging, making it difficult to support complex cross-model queries with acceptable performance ~\cite{khine2019review, DBLP:journals/pvldb/KiehnSGPWWPR22, lajam2022revisiting}

We are not aware of a commercially available polyglot system that combines graph and time-series DBMS. We therefore included a research prototype developed at Leipzig University, TimeTravelDB, which is designed to store both graph and time-series data natively and consists of a new query language, TimeTravel Query Language (TTQL), to manage models and their combinations.

An example query for such a polyglot approach is presented in \cite{rep_queries}.

  \begin{table*}[t]
\centering
\caption{Overview of System Architectures for Graph + Time Series}
\renewcommand{\arraystretch}{1.4}
\begin{tabular}{@{}p{3.3cm}p{4.1cm}p{4.9cm}p{4.0cm} c@{}}
\hline
\textbf{Category} & \textbf{Key Benefits} & \textbf{Key Challenges} & \textbf{Example systems} \\
\hline
\textbf{Single-Model}& leveraging mature DBMS. ACID across models since one engine.& optimized for at most one of the models (graph or time series), no support for hybrid queries & Neo4j~\cite{neo4j}, InfluxDB~\cite{influxdb}, Memgraph~\cite{memgraph}, MongoDB~\cite{mongodb}, Aion~\cite{theodorakis2024aion}, ClockG~\cite{massri2022clock} \\
\hline
\textbf{Extended Single-Model} & like for SM plus dedicated support for  graph/time series data& extensions must align with the core model; potential performance overhead&  PostgreSQL~\cite{postgres}, Redis~\cite{redis},  Couchbase~\cite{couchbase}, MariaDB~\cite{mariadb}, Cassandra~\cite{cassandra}, Oracle~\cite{oracle}   \\
\hline
\textbf{Polyglot Persistence} & native support of both data models. %scaling issues are pushed to databases. 
& high integration complexity, high latency for queries on both kinds of data, no cross-model transactions, no mature implementation & TimeTravelDB prototype~\cite{timetraveldb} \\
\hline

\textbf{ MMDB}  & single engine supports multiple models, ACID across models  & complex to implement; restricted maturity of current implementations; graph/time series processing less optimized than in native DBMS & ArangoDB~\cite{arangodb}, OrientDB~\cite{orientdb}, ArcadeDB~\cite{arcadedb}, BangDB~\cite{bangdb}, SurrealDB~\cite{surrealdb}\\

\hline
\end{tabular}
\label{tab:architecture_comparison}
\end{table*}
 \subsection{Multi-model database (MM)}
\label{subsec:mmn} 

Multi-model DBMS (MMDB) \cite{zhang2019unibench} provides a unified platform for managing data across multiple models and data types, such as documents, graphs, relational tables, and time series, within a single system. While polyglot persistence relies on separate, specialized data stores for different data models, an MMDB integrates these models, offering a more flexible and consistent approach to data management with fewer data synchronization issues \cite{lu2019multi}. In contrast to E-SM systems, all supported data models are treated equally as 'first-class citizens', ensuring consistent performance and capabilities across all models~\cite{guo2024multi, lu2019multi}. MMDB systems use a single engine to natively handle all supported data models without relying on additional layers, modules, plug-ins, or middleware~\cite{arangodbdoc, guo2024multi}. While there can be support for multiple query languages for different models, there should ideally be a 
 unified query language that seamlessly manages and queries all supported data models. This would support complex multi-model queries without the need to switch between different query languages~\cite{lu2019multi}. For example, ArcadeDB supports Cypher and Gremlin for graph queries in addition to SQL-like functionality. By contrast,  ArangoDB provides a complex unified query language called AQL ~\cite{mavrogiorgos2021comparative}.

While MMDB systems appear promising, they are challenging to implement with robust functionality and performance, depending on the extent to which the system can integrate data from different models. This is also complicated by the 
%separate or  Native multi-model databases, built from the ground up to support multiple data models, face the daunting task of initial development and implementation. The 
lack of established standards and best practices for integrating and optimizing various data models~\cite{lu2019multi,koupil2023mm,lu2018udbms}. We will see that the current MMDBs remain quite restricted in meeting our requirements. The close integration of 
graph and time series data with a uniform query language, as envisioned in ~\cite{ammar2025towards}, could be supported by an advanced MMDB but has not yet been implemented.

%within a single database system makes development time-consuming and resource-intensive. Furthermore, having a single query language for multiple data models may lead to a more complex syntax, such as AQL, the query language of Arangodb~\cite{mavrogiorgos2021comparative}

% \subsection{Unified hybrid model }
 
% This degree of unification is the one desired, achieved by treating graphs and time series as a single, composite data model with its own query operators and combined analysis that considers both dimensions (structural and temporal) simultaneously. Unlike the previous categories, we do not provide an architecture diagram for this category because no system neither industrial nor research prototype has yet achieved a complete implementation satisfying all requirements. Our paper provides a further explanation of this vision.

\section{Comparative evaluation}
\label{sec:results}
In our evaluation, we compare 15 commercial data management systems and five research prototypes of the four integration architectures w.r.t. the introduced requirements (\Cref{tab:modeling}) and implementation characteristics (\Cref{tab:charac})\footnote{For space reasons, we moved some requirements, such as support for OLTP and OLAP, to \Cref{tab:charac}}. While it is not possible to consider all relevant systems, we identified representative solutions for each of the four categories that at least partially meet the requirements for processing both graph and time-series data. In the following, we discuss the results separately for the four architectures.

In \Cref{tab:modeling}, we classify systems with native stream processing as fully supporting \textit{streaming} (e.g., with real-time ingestion and analytical pipelines), systems that rely on external stream-processing engines via connectors as partially supporting streaming, and systems without explicit streaming capabilities beyond batch operations as not supporting streaming.
Regarding \textit{bidirectional model transformation}, we classify 
a system supporting both directions (from graph/ts → hybrid and back) as full support, and for support of only one direction (e.g., importing hybrid data but not exporting it or exporting only a single model) as partial support.  
In \Cref{tab:charac}, full support of \textit{Graph-TS indexing} requires indexing for both data types and composite indexing. In contrast, partial support is indicated when only one of the data types is indexed or composite Graph-TS indexing is missing.
Several industrial systems provide multiple editions (e.g., community and enterprise). In our comparison, a system is considered open source if it offers an OSI-licensed community edition and if the capabilities reported in our tables are available in that community edition without requiring a paid license. If the reported capabilities rely on enterprise-only features, the system is not considered open source in our evaluation.
For example, Neo4j provides both a community and an enterprise edition. While most graph query operations are supported in the community edition, scalability is only available in the enterprise edition. Accordingly, we consider Neo4j to be open source but not scalable. In contrast, the Oracle database is distributed under a commercial license. Although limited free editions exist, they are not open source. Therefore, Oracle is treated as a closed-source system, and the reported capabilities correspond to its enterprise offering.

\begin{table*}[htbp]
\centering
\small
\caption{Modeling and Query Capabilities across Systems. Legend: \Yes = supported; \Partial = workaround or via extension/plugin; \No = not supported. RP = research prototype. KV = key value.
Symbols: \Yes~= native/first-class; \Partial~= via extension or workaround; \No~= unsupported.}
\label{tab:modeling}
\renewcommand{\arraystretch}{0.9}

% Changes applied:
% (1) Swapped column order: Native TS is now before Temporal evolution
% (2) Added Streaming column right after Hybrid model

\begin{tabular}{@{}p{0.8cm}p{1.7cm}p{2.9cm}c c c c c p{2.0cm} c p{3.1cm} c c c@{}}
\toprule
\begin{sideways}\textbf{Architecture}\end{sideways} &
\begin{sideways}\textbf{System}\end{sideways} &
\begin{sideways}\textbf{Primary Model}\end{sideways} &
\begin{sideways}\textbf{Native PG}\end{sideways} &
\begin{sideways}\textbf{Native TS}\end{sideways} &
\begin{sideways}\textbf{Temporal evolution}\end{sideways} &
\begin{sideways}\textbf{Hybrid model}\end{sideways} &
\begin{sideways}\textbf{Streaming}\end{sideways} &
\begin{sideways}\textbf{Query Language}\end{sideways} &
\begin{sideways}\textbf{Graph Pattern}\end{sideways} &
\begin{sideways}\textbf{TS analysis}\end{sideways} &
\begin{sideways}\textbf{Cross-model Query}\end{sideways}  &
\begin{sideways}\textbf{Bidirectional}\end{sideways}
\begin{sideways}\textbf{transformation}\end{sideways}&
\begin{sideways}\textbf{Acid cross-model}\end{sideways}\\
\midrule

SM & InfluxDB & Time Series & \No & \Yes & \No & \No & \Yes & InfluxQL & \No &
\makecell[l]{Aggregations,\\grouping, join, etc} & \No & \No & \No\\

& Memgraph & Property Graph & \Yes & \No & \No & \No & \Yes & openCypher & \Yes &
\No & \No & \No & \No\\

& MongoDB & Document & \Partial & \Partial & \No & \Partial & \Partial & MQL & \Partial &
\makecell[l]{\ttfunc{\$bucket}, \ttfunc{\$window},\\ etc.\ in agg.\ pipeline} & \Partial & \No & \Partial \\

& Neo4j & Property Graph & \Yes & \No & \No & \No & \Partial & Cypher & \Yes &
\No & \No & \No & \No\\
\midrule

SM (RP) & Aion & Temporal Graph & \Yes & \No & \Yes & \No & \No & Temporal Cypher & \Yes &
\No & \No & \No & \No\\

& AeonG & Temporal Graph & \Yes & \No & \Yes & \No & \No & Temporal Cypher & \Yes &
\No & \No & \No & \No\\

& Bollen et al. & Property graph with time series & \Yes & \Partial & \No & \Yes & \No & GQL-TS & \Yes &
Windows, aggregations and TS pattern & \Yes & \No & \No\\

& Clock-G & Temporal Graph & \Yes & \No & \Yes & \No & \No & T-Cypher & \Yes &
\No & \No & \No & \No\\
\midrule

E-SM & Couchbase & Document & \No &
\href{https://www.couchbase.com/blog/introducing-couchbase-time-series/}{\Partial} &
\No & \No & \Partial & SQL++ &
\href{https://www.couchbase.com/blog/query-graph-recursive-cte/}{\Partial} &
\href{https://docs.couchbase.com/server/current/n1ql/n1ql-language-reference/timeseries.html}{\ttfunc{\_TIMESERIES()}} &
\No & \No  & \Partial \\

& Cassandra* & Wide-column store & \Partial &
\href{https://cloudinfrastructureservices.co.uk/cassandra-data-modeling-patterns-time-series-best-practices/#:~:text=A%20sequence%20of%20data%20points,Wide%20Row%20Pattern}{\Partial} &
\No & \No & \Partial & CQL &
\No &
\href{https://www.mongodb.com/docs/manual/core/timeseries/timeseries-aggregations-operators/}{\makecell[l]{Aggregation functions\\ \ttfunc{\$merge}, \ttfunc{\$out}, \ttfunc{\$group}, etc.}} &
\No & \No & \No\\

& MariaDB & Relational &
\href{https://mariadb.com/docs/server/server-usage/storage-engines/oqgraph-storage-engine}{\Partial} &
\href{https://mariadb.com/resources/blog/analysis-of-financial-time-series-data-using-mariadb-columnstore/}{\Partial} &
\href{https://mariadb.com/docs/server/reference/sql-structure/temporal-tables/bitemporal-tables}{\Partial} &
\No & \Partial & SQL &
\Partial &
Aggregations, statistical functions over window &
\Partial & \No & \No\\

& Oracle & Relational & \Yes & \Partial & \Partial & \Partial & \Yes & SQL, PGQL & \Yes &
\href{https://docs.oracle.com/en/database/oracle/machine-learning/oml4sql/23/mlsql/time-series.html}{ML, TS regression, forecasting, statistics} &
\Yes & \No &\Yes \\

& PostgreSQL* & Relational &
\href{https://age.apache.org/}{\Partial} &
\href{https://www.tigerdata.com/}{\Partial} &
\No & \Partial & \Partial & SQL &
\Partial &
\href{https://maddevs.io/writeups/time-series-data-management-with-timescaledb/}{\makecell[l]{Timescale: GROUP BY,\\ time filters, aggregations}} &
\Yes & \No & \Partial \\

& Redis* & Key-value &
\href{https://redis.io/docs/latest/operate/oss_and_stack/stack-with-enterprise/deprecated-features/graph/}{\Partial} &
\href{https://redis.io/docs/latest/operate/oss_and_stack/stack-with-enterprise/timeseries/}{\Partial} &
\No & \Partial & \Partial & Commands / Cypher (graph) &
\Partial &
\href{https://redis.io/docs/latest/develop/data-types/timeseries/}{\makecell[l]{RedisTS: TS.GET,\\ TS.RANGE, etc.}} &
\Partial & \No & \No\\
\midrule

Polyglot (RP) & TimeTravelDB & TTPGM (1) & \Yes & \Yes & \Yes & \Yes & \No & TTQL (Extended Cypher) & \Yes &
\makecell[l]{Native (FROM, TO,\\ SHALLOW), aggregations} &
\Yes & \Partial & \No  \\
\midrule

MM & ArangoDB & Document + Graph + KV & \Yes & \No & \No & \No & \Partial & AQL & \Yes &
\No & \Partial & \No & \Yes \\

& ArcadeDB & \makecell[l]{Document + Graph +\\ Time Series + KV} & \Yes & \Partial & \No & \Partial & \Partial & SQL-like & \Yes &
\No & \Partial  & \No & \Partial\\

& BangDB & Document + Graph + Time Series & \Yes & \Yes & \No & \Partial & \Yes & \makecell[l]{C++,\\ SQL-like} & \Yes &
\href{https://bangdb.com/products/bangdb/stream-processing}{\makecell[l]{Join, filter, aggregations,\\ sliding window, etc.}} &
\Yes  & \No & \Partial\\

& OrientDB & Document + Graph + KV & \Yes & \No & \No & \No & \Partial & SQL-like & \Yes &
\No & \Partial  & \No & \Yes\\

& SurrealDB & \makecell[l]{Document + Graph +\\ Time Series (+others)} & \Yes & \Yes & \No & \No & \No & SurrealQL &
\href{https://surrealdb.com/docs/surrealdb/models/graph}{\Yes} &
\href{https://surrealdb.com/docs/surrealdb/models/time-series}{\makecell[l]{Aggregations,\\ GROUP BY, etc.}} &
\Partial  & \No & \Partial\\

\bottomrule
\end{tabular}

\footnotesize \textbf{Notes:} Asterisks (*) indicate plugin-based capabilities. RedisGraph EOL on \textbf{Jan 31, 2025}. (1) TimeTravel Property Graph Model
\end{table*}

\begin{table*}[htbp]
\centering
\begingroup
\small
\caption{Implementation characteristics. Legend: \Yes = native/first-class. \Partial = partial or via extension/plugin. \No = not supported. CQ = Continuous query. CQN = continuous query notification. Scalability \& Availability codes: S = Sharding; R = Replication; SR = Sharding + Replication. KV=key-value}
\label{tab:charac}
\renewcommand{\arraystretch}{0.9}

% Graph-TS Index moved right after Storage Structure
\begin{tabular}{@{}p{1.2cm}p{1.6cm}p{2.7cm}c c c c c c c p{2.4cm}@{}}
\toprule
\begin{sideways}\textbf{Architecture}\end{sideways}&
\begin{sideways}\textbf{System}\end{sideways} &
\begin{sideways}\textbf{Storage}\end{sideways} \begin{sideways}\textbf{Structure}\end{sideways}&
\begin{sideways}\textbf{Graph-TS Index}\end{sideways} &
\begin{sideways}\textbf{Disk/In-Memory}\end{sideways}&
\begin{sideways}\textbf{OLTP}\end{sideways}&
\begin{sideways}\textbf{OLAP}\end{sideways}&
\begin{sideways}\textbf{Active DB}\end{sideways} &
\begin{sideways}\textbf{Scalability \&}\end{sideways}
\begin{sideways}\textbf{Availability}\end{sideways}
&
\begin{sideways}\textbf{Retention Policy}\end{sideways} &
\begin{sideways}\textbf{License}\end{sideways}\\
\midrule

SM & InfluxDB
& \href{https://docs.influxdata.com/influxdb/v2/reference/internals/storage-engine}{Time-Structured Merge Tree (TSM)}
& \Partial
& \href{https://www.alibabacloud.com/blog/594732}{Disk}
& \Partial
& \Yes
& \href{https://www.alibabacloud.com/blog/594732}{\Yes~CQ}
& R\,(1)
& \href{https://docs.influxdata.com/influxdb/v2/reference/internals/data-retention/}{\Yes}
& MIT (open source) \\

& Memgraph
& In-memory (skip lists)
& \Partial
& In-memory
& \Yes
& \Partial
& \href{https://memgraph.com/docs/fundamentals/triggers}{\Partial~Trigger}
& R
& \href{https://memgraph.com/docs/querying/time-to-live}{\Partial}
& BSL 1.1 (source-available) \\

& MongoDB
& \href{https://groups.google.com/g/mongodb-dev/c/pe99ieBlW4I}{B-tree (WiredTiger)}
& \Partial
& Disk
& \Yes
& \No
& \href{https://medium.com/@hirunagrad/database-triggers-how-to-setup-mongodb-app-services-function-triggers-74d800ab6b91}{\Partial~Trigger}
& \href{https://www.geeksforgeeks.org/mongodb/mongodb-replication-and-sharding}{SR}
& \href{https://www.mongodb.com/community/forums/t/gdpr-retention-policies-deletion/105547}{\Partial}
& SSPL (source-available) \\

& Neo4j
& Fixed-size record store
& \href{https://neo4j.com/docs/cypher-manual/current/values-and-types/temporal/#cypher-temporal-index}{\Partial}
& \href{https://memgraph.com/blog/neo4j-vs-memgraph}{Disk}
& \Yes
& \Partial
& \Partial~Trigger
& \No\,(1)
& \No
& GPL v3 (Community) \\

\midrule

SM (RP) & Aion
& TimeStore + LineageStore
& \Partial
& Both
& \Yes
& \Yes
& \Partial~Event listener
& SR
& \No
& Open source \\

& AeonG
& Current store (MVCC) + historical KV (anchor+delta)
& \Partial
& Disk
& \Yes
& \Yes
& \No
& S\,(4)
& \Yes
& RP \\

& Bollen et al.
& Fixed-size record store (Neo4j)
& \No
& Disk
& \Yes
& \No
& \No
& \No
& \No
& RP \\

& Clock-G
& $\delta$-Copy+Log (Cassandra backend)
& \No
& Disk
& \Partial
& \Yes
& \No
& \No
& \No
& RP \\

\midrule

E-SM & Couchbase
& \href{https://www.vldb.org/pvldb/vol15/p3496-lakshman.pdf}{B-tree (Couchstore/Magma)}
& \No
& Both
& \Yes
& \href{https://docs.couchbase.com/server/current/learn/services-and-indexes/services/analytics-service.html}{\Yes}
& \href{https://www.couchbase.com/blog/eventing-data-consolidation-in-couchbase}{\Partial~Eventing service}
& SR
& \href{https://docs.couchbase.com/server/current/learn/data/expiration.html}{\Partial}
& BSL 1.1 (source-available) \\

& Cassandra*
& \href{https://cassandra.apache.org/doc/latest/cassandra/architecture/storage-engine.html}{Log-structured merge-tree (SSTables)}
& \No
& Disk
& \Yes
& \href{https://medium.com/doublecloud-insights/clickhouse-vs-cassandra-an-in-depth-comparison-for-data-architects-affd5b4facea}{\Partial}
& \Partial~Java trigger
& SR
& \href{https://docs.datastax.com/en/cql-oss/3.3/cql/cql_using/useExpire.html}{\Partial}
& Apache 2.0 \\

& MariaDB
& \href{https://dev.mysql.com/doc/refman/8.4/en/innodb-physical-structure.html}{Row store (InnoDB)\,(2)}
& \No
& Both
& \Yes
& \href{https://mariadb.com/docs/columnstore/architecture/columnstore-storage-architecture}{\Yes}
& \href{https://www.mariadbtutorial.com/mariadb-triggers}{\Yes~Trigger}
& SR
& \No
& GPL v2 \\

& Oracle
& row store table\,(5)
& \Partial
& Both
& \Yes
& \Yes
& \Yes~Trigger/CQN
& SR
& \href{https://www.oracle.com/assets/automatic-data-optimization-wp-12c-1896120.pdf}{\Partial}
& Commercial \\

& PostgreSQL*
& \href{https://dba.stackexchange.com/questions/337831/why-do-databases-use-heap-to-store-table-data-while-other-structure-like-b-tree}{ Row store (B-tree / HEAP)}
& \Partial
& Disk
& \Yes
& \Partial
& \Yes~Trigger
& S\,(3)
& \href{https://www.tigerdata.com/learn/what-is-data-retention-policy}{\Yes}
& PostgreSQL License \\

& Redis*
& \href{https://redis.io/docs/latest/operate/oss_and_stack/management/persistence/}{In-memory (hash + skip list)}
& \Partial
& In-memory 
& \Yes
& \Partial
& \href{https://redis.io/blog/introducing-triggers-and-functions}{\Yes~Trigger}
& S
& \href{https://redis.io/docs/latest/operate/oss_and_stack/stack-with-enterprise/timeseries}{\Yes}
& BSD 3-Clause \\

\midrule

Polyglot (RP) & TimeTravelDB
& Record store + columnar
& \Partial
& Disk
& \Yes
& \No
& \Partial~Trigger + CQ
& S
& \Yes
& Open source \\

\midrule

MM & ArangoDB
& \href{https://docs.arangodb.com/3.12/components/arangodb-server/storage-engine/}{LSM-tree (RocksDB)}
& \Partial
& Disk
& \Yes
& \Partial
& \href{https://docs.arangodb.com/stable/develop/integrations/spring-data-arangodb/reference-version-4/mapping/events/}{\Partial~Spring event}
& SR
& \No
& BSL 1.1 \\

& ArcadeDB
& \href{https://docs.arcadedb.com/#indexes}{LSM-tree}
& \Partial
& Disk
& \Yes
& \Partial
& \href{https://docs.arcadedb.com/#java-api}{\Partial~Java listener}
& R
& \No
& Apache 2.0 \\

& BangDB
& \href{https://docs.bangdb.com/access-methods?q=data+structure}{B-tree + Hash}
& \Partial
& Both
& \Yes
& \Yes
& \No
& SR
& \No
& BSD 3-Clause \\

& OrientDB
& \href{https://www.tutorialspoint.com/orientdb/orientdb_quick_guide.htm}{Paginated Local Storage}
& \Partial
& Both
& \Yes
& \Partial
& \href{https://docs.huihoo.com/orientdb/1.7.8/orientdb.wiki/Hook.html}{\Yes~Trigger}
& SR
& \No
& Apache 2.0 (Community) \\

& SurrealDB
& \href{https://surrealdb.com/learn/fundamentals/performance/deployment-storage}{LSM tree (RocksDB)}
& \Partial
& Both
& \Yes
& \Partial
& \href{https://surrealdb.com/docs/surrealdb/models/time-series}{\Yes~Live queries}
& SR
& \Partial
& BSL 1.1 (source-available) \\

\bottomrule
\end{tabular}
\endgroup

\footnotesize \textbf{Notes:} (1) Entreprise = HR. (2) ColumnStore engine available for time series. (3) H with Citus. (4) Distributed variant AeonG-D. (5) In-memory column store available
\end{table*}

\subsection{Single model systems result}
For systems of type \textbf{SM}, we include native property graph DBMS (Neo4j, Memgraph), native time series DBMS (InfluxDB), and a document store (MongoDB). Furthermore, we consider three research prototypes for temporal graph data (Aion, AeonG, Clock-G) and one prototype for property graphs with time-series support (Bollen et al.~\cite{bollen2024managing}).  
%Single-model (SM) systems fall into four families: (i) \emph{graph DBMSs} that natively support the PG model and encode time series via workarounds, (ii) \emph{time series DBMSs} whose data model can encode graph structure inside the TS schema, and (iii) \emph{document stores} that emulate both PG and TS with JSON along with some data oriented operators. 
%Across all four, the integration is inherently \emph{asymmetric}: one model is first-class, the other is an encoded add-on. As a consequence, it's hard for these systems to provide deep cross-model optimization and cross-model transactional guarantees. 
As expected and indicated in \Cref{tab:modeling}, no SM system fully supports both property graphs (PG) and time series (TS). Cross-model queries and ACID transactions are largely unsupported, thereby leaving essential requirements unsatisfied.   
%and TS natively; at best, they offer \emph{partial} support for the non-native model (via specific operators or added engines), and many offer none. Hybrid capabilities (i.e., explicit linkages between graph and time-series entities), hybrid operators, and hybrid updates are largely absent, reflecting the goal of a single-model design.

The considered  \textit{graph systems}  Neo4J and Memgraph have no built-in support for time series. One could model time series by property values. Still, with simple property values, this would require a large number of $2n$ property values for a time series of length $n$ (one property value per event, for the timestamp and the value). In Neo4j, the entire time series can also be represented as a graph with a chain of event nodes or using the GraphAware Timetree plugin~\cite{graphawaretimetree}.  This encoding enables hybrid queries (pattern matching constrained by temporal predicates or aggregations), but at the expense of verbose, complex query formulations and, consequently, poor usability. 
The research prototype by Bollen et al. includes support for time series and cross-model queries in its query language GQL-TS  and represents time series as a linked list of nodes~\cite{bollen2024managing}.  
 Neo4J,  Memgraph, and prototype ~\cite{bollen2024managing} focus on OLTP, but OLAP-like queries are also possible. The in-memory system  Memgraph has support for streaming (graph) data.  

The \textit{temporal graph prototypes} keep historical graph data and can thus provide support of evolution and snapshot analysis, 
but do not natively support time-series analysis or hybrid queries. The query language is usually based on Neo4J's Cypher, extended with temporal predicates, e.g., to derive snapshots at a specific point in time. Both OLTP and OLAP queries are typically supported. The prototype AeonG\footnote{\url{ https://github.com/hououou/AeonG}} is under BSL1.1 licence, which means it's closed source, while Aion\footnote{\url{ https://github.com/aeon-toolkit/aeon}} is open-source. No code implementation is available for Clock-G and  Bollen et al.'s method. 

The \textit{time-series DBMS} InfluxDB has no built-in support for storing and analyzing graph data or cross-model queries. Time-series events (measurements) can be enriched by metadata that can be used to store node states or relationships between nodes.  Hence, there are ways for users to represent and analyze time-evolving graph data, but not so much for structural graphs and complex graph traversals and analyses.  InfluxDB is an OLAP system for time-series data and supports streaming data. 

    The other relational and NoSQL SM systems provide no native support for graph and time-series data. Still, it is possible to represent and analyze both types of data using the underlying data model and query language. In the considered \textit{document DBMS}, MongoDB graph nodes and edges can be represented by documents of typed collections, and there is an operator \textit{\$graphLookup} for simple graph analysis.  TS data can be stored in time-series collections, introduced in MongoDB 5.0, and analyzed with aggregation operators.     
    %They are created with a \textit{timeField} (and an optional \textit{metaField}) and stored in an internal “\textit{buckets}” collection using a compressed columnar layout, clustered by time and metadata for efficient writes and range scans. 
    There is support for compressing time-series data and for a TTL (time-to-live) policy for data retention. 
    Hence, one can achieve some basic support for both data types and mixed queries, but without advanced or optimized capabilities.
 %   For graph data, MongoDB models nodes/edges as ordinary JSON documents. Recursive traversal is available via the  aggregation stage. Still, MongoDB is not a specialized graph database. One solution is to create two collections: one for $nodes$ and one for $edges$. The time-series data for each is stored in a native MongoDB time-series collection named metrics. This yields good engineering flexibility but weak model semantics (no native graph constraints or temporal operators) and limited cross-model optimization.

The implementation characteristics (\Cref{tab:charac}) are quite heterogeneous, with a focus on disk storage and partial support for sharding and replication.  
%storage engines are tuned for one modality: adjacency-list layouts for PG, LSM/TSM-style layouts for time-series, and B$^{+}$-tree row stores for relational systems. These choices maximize the native workload (e.g., traversals or time-window scans) but impede cost-based optimization across PG and TS operators. 
%Hence, SM systems achieve only \emph{shallow} integration: expressive for their home model but reliant on verbose encodings and application logic for the other.

%\textbf{Summary.}

%SM systems provide high maturity for a single model but only shallow integration with the other model. Encodings introduce (i) verbose queries, (ii) absence of native snapshot/interval semantics for graph patterns, (iii) limited cost-based optimization across LPG and TS operators, and (iv) no cross-model transactional guarantees. These observations are consistent with Table~\ref{tab:modeling}–\ref{tab:charac}: SM rows show native support on their home model, partial/none on the other, and limited (or absent) cross-query/hybrid-update capabilities.
%\begin{figure*}[t] % spans both columns
%  \centering
%  \begin{subfigure}[b]{0.49\textwidth}
%    \includegraphics[width=\linewidth]{figures/char_heatmap_v2 (1).png}
%    \subcaption{Coverage of characteristics across systems}
%    \label{fig:left}
%  \end{subfigure}\hfill
%  \begin{subfigure}[b]{0.49\textwidth}
%    \includegraphics[width=\linewidth]{figures/requirements_heatmap_v2 (1).png}
%    \subcaption{Coverage of requirements across systems} 
%    \label{fig:right}
%  \end{subfigure}
%  \label{fig:two-wide}
%\end{figure*}

\begin{figure}
    \centering
    \includegraphics[width=1\linewidth]{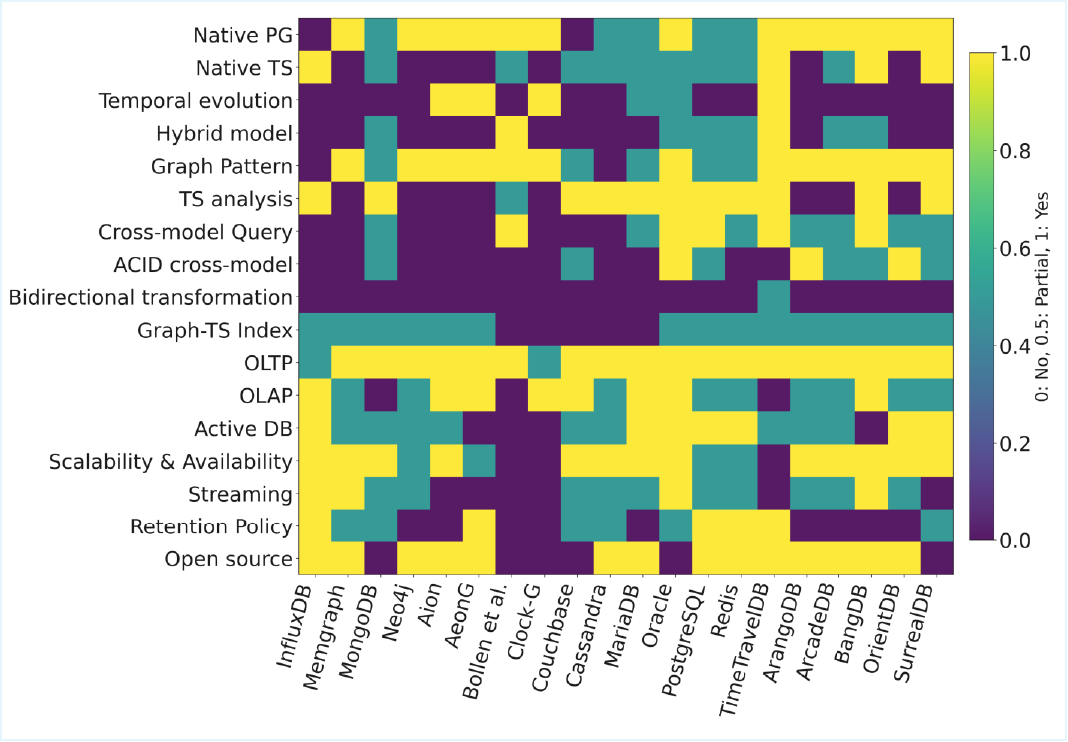}
    \caption{Coverage of requirements and characteristics across systems}
    \label{fig:heatmap}
\end{figure}
\vspace{2cm}
\subsection{Extended single-model (E-SM) results}
For type  E-SM, we consider three relational DBMS (Oracle, PostgreSQL, MariaDB) and three NoSQL systems (Redis, Couchbase, Cassandra).
%E-SM systems run multiple paradigms inside a single engine via extensions or embedded services. Representative examples in our study are \textit{Couchbase} (SQL++ with \_TIMESERIES()), \textit{Cassandra}(Datastax graph extension /aggregation patterns), \textit{MariaDB} (UDFs/aggregation patterns), \textit{PostgreSQL*} (TimescaleDB + AGE), and \textit{Redis*} (RedisTimeSeries + Graph, now EOL). 
These systems primarily support graph and time-series data via extensions (requirement D1), partially support cross-model queries (Q1), and partially support ACID across data types (Q3), but still lack a unified hybrid model (D2) to integrate both data types within a single data structure. There is support for streaming data (D4), but not much for temporal evolution (D3). In E-SM systems, scalability guarantees provided by the core engine do not necessarily extend to graph and time series extensions, and vice versa. Therefore, in this survey, we consider a system to fully support scalability only if both the core engine and its extension are scalable.

%in a ( with a close with first-class time-series PG semantics (D2–D3 are partial). Because execution is centralized, these systems can perform cross-model joins in one statement (Q2) and support hybrid ingestion pipelines (Q1). \emph{partial} ACID across models (Q3) is possible when both sides live under the same transaction manager (e.g., Timescale tables + AGE relations in one Postgres instance). However, extension boundaries and heterogeneous storage paths often limit the depth of hybrid support. Temporal indexing and retention tend to be strong when the TS extension is mature (T-index/TTL, D4). Our results show that E-SM scores well in Q1, with limited performance in D2/D3 and Q3, reflecting the “trade-off "shared runtime, heterogeneous semantics.
%Relational DBMSs store vertices and edges as tables, and connections as foreign keys (e.g., vertex source and target ids inside the edge table). Some relational databases offer more flexibility in supporting data. 
The relational DBMS Oracle provides strong support for property graphs and for analytic analysis of time-series data stored in relational tables. It can also manage the temporal evolution of both graph and time-series data with support for historical querying  (D3). The graph capabilities of Oracle Database 23ai
%enabling graph queries to be combined with relational and JSON data within a single engine (Q1). Oracle Graph provides an
include an extensive library of graph algorithms (more than 80~\cite {oracle_graph}). Time-series analysis is supported through SQL functions~\cite{oracle_timeseries} (e.g., \texttt{TIME\_BUCKET}, \texttt{AGO}, \texttt{TODATE}) and Oracle Machine Learning for SQL for time-series forecasting~\cite{oracle_timeseries_ml}. 
%As both graph and time-series data reside in the same transactional engine, cross-model operations can rely on standard ACID guarantees (Q3), and 
Oracle can thus serve both OLTP and OLAP workloads.
%without requiring a separate analytical system for many hybrid scenarios (Q4). 
Streaming support is possible via Oracle   Cloud Infrastructure and a Kafka-compatible  event streaming service ~\cite{kafka}
%with features such as message replay and ordered partitions~\cite{kafka}, 
which can be used for real-time ingestion of graph and time-series data into Oracle.

PostgreSQL is a relational database that provides extensions for time-series data via TimescaleDB~\cite{timescaldb} and for graph data via Apache AGE~\cite{apache-age}. This also includes support for cross-model queries (Q1) and  \emph{partially} ACID across models (Q3) provided that Timescale and AGE tables are managed by the same  Postgres instance. 
However, extension boundaries and heterogeneous storage paths often limit the depth of hybrid support. PostgreSQL supports only partial data streaming (D4) via its TimescaleDB extension, which enables real-time data ingestion and querying. 

The relational DBMS MariaDB supports multiple storage engines on the same server, including engine \emph{OQGRAPH} to handle complex data in a graph-like structure~\cite{OQGRAPH_mariadb}, as well as bi-temporal tables for validity and transaction time~\cite{bui2022bitemporal}. There is no dedicated storage adaptation for time-series data, but such data can be handled by the \emph{ColumnStore} engine, which is optimized for analytic queries~\cite{ts_mariadb}. 
Support for cross-model queries is low. They require an ETL-like step to communicate between engines for query processing: exporting results from one engine and importing them into another for a combined query execution. 

%Although expressive in SQL (partial Q1), the approach lacks a unified representation across the two data models (Q2), leading to manual feature orchestration and non-trivial maintenance of consistency between graph and time-series data (Q3).  
% \color{red}
Redis is an open-source, in-memory key-value store that includes extensions for time-series data (RedisTimeSeries~\cite{redis_ts}) and graph data (RedisGraph~\cite{redis_graph}, which uses the OpenCypher query language), although RedisGraph is no longer supported. Each extension provides native query capabilities for its respective data model, 
%including retrieving data within a time range, time-series aggregations, downsampling of time-series data, and pattern matching and graph path-finding algorithms (Q4: OLTP and lightweight OLAP). However, it does not provide 
but there is no built-in support for cross-model queries. 
%Therefore, it requires issuing separate queries to the respective extension modules, after which the results can be merged using Redis functions.  
Redis provides partial streaming support via the Redis Streams extension~\cite {redis_streams}. Data retention support (P5) is provided only through RedisTimeSeries (P5). 
%by enabling ordered even ingestion, but no advanced stream processing features such as integrated analytics pipelines (D4).    

 The distributed document store Couchbase has no native support for graph and time-series data. Still, it can store these data types in JSON documents and provides optimizations and dedicated analysis capabilities for time series data ~\cite{couchbase_timeseries}. Its SQL-like query language, SQL++, also supports recursive queries, e.g., for reachability queries and graph traversals.
 % is adatabase that exposes an SQL-like query language (SQL++). Couchbase doesn't provide native support for storing graph data, but it can be stored as a JSON document, just as it does for time series data storage~\cite{couchbase_timeseries} but with efficient storage to handle the large volume of time series data (D1). It, however, provides a recursive querying pattern in its query language (e.g., for reachability/ path-style queries) for graph traversal. Time series analysis is supported through time-oriented querying/aggregations with a time range (Q2-Q4). 
There is no  built-in support for  cross-model queries so users  have to implement the combination  
%however, it could be implemented by joining 
of graph documents with time-series documents. 
Streaming is supported only via connectors to an external streaming system.
 %\color{black}
 
The primary reason for including Cassandra is its high scalability as a wide-column store, due to its decentralized storage architecture~\cite{lakshman2010cassandra}, and its use as backend storage for time-series databases, such as KairosDB~\cite{kairos}, and for graph databases, such as JanusGraph. Both can be used within a single Cassandra engine.

\subsection{Polyglot persistence}
For polyglot persistence, we consider the research prototype TimeTravelDB (TTDB) \cite{ammar2025towards} due to a lack of a suitable commercial system. 
%Polyglot deployments comprise two (or more) specialized engines behind a mediator/adapter. This architecture maximizes per-model fidelity (D1) and leverages native operators/indexes in each store, but lacks a unified logical model (D2). Cross-model querying (Q1) is achieved via query decomposition and runtime joins across results. %Operator pushdown is limited to per-store fragments, resulting in weak global cost-based optimization. 
%Hybrid ingestion/timeliness (Q3) depends on the mediator’s change data capture (CDC) and scheduling. Without atomic multi-store commits, ACID across models is typically unavailable (Q3). On the upside, the polyglot inherits mature OLTP/OLAP per-store (D4), temporal indexing and TSDB retention, and horizontal scale (P6) through the native sharding/replication of each system. 
%{TimeTravelDB} (TTDB), our research prototype~\cite{ammar2025towards}, is a multistore system developed in our department. 
It combines the graph DBMS Neo4j and the time-series extension TimescaleDB of PostgresSQL. 
In its middleware layer it supports an extended temporal property graph data model called TTPGM (Time Travel Property Graph Model) (D1)
where graph nodes can possess time-series properties. A UUID (universally unique identifier) references time-series data within graph elements. This is the “hybrid link” that maintains consistency between the two separated databases (D2).
%following the  (TTPGM), a temporal property graph data model (D3) that extends graph entities with time-series data. Thus, properties can be static or dynamic when related to time-series data. 

For query processing an extended Cypher query language called TTQL (TimeTravel query language) supports temporal 
%clauses \emph{From ... To ...} that support 
range queries and aggregations on time-series data~\cite{timetraveldb} (Q1). 
%A thin layer: API → TTQL parser → query processor → storage engine, 
Query processing entails splitting a user query into a Cypher component (graph) and an SQL component (time-series), executing both components, and merging the results.  TimeTravelDB derives its OLTP capabilities from its underlying systems but remains weak in OLAP, providing only pattern matching with time-series constraints.
The TTDB prototype supports temporality through a validity property per graph element and temporal range queries (keywords {FROM, TO}) (D3). %For any time $t$ in the graph's time domain $T$, the predicate returns \textbf{true} if a given vertex or edge "exists" at $t$, and \textbf{false} otherwise. Practically, this means 

%\vspace{2cm}
\subsection{Multi-model systems (MM)}
We consider five multi-model implementations (  ArangoDB, OrientDB, ArcadeDB, BangDB, SurrealDB). 
%MM engines integrate multiple models natively into a single storage/runtime, with a single optimizer and transaction layer. 
They all provide native support for documents and graphs, but not all support time series, as these can be stored in other data structures, such as ordered lists and key-value maps. %In our set, \textit{ArangoDB}, \textit{OrientDB}, \textit{ArcadeDB}, \textit{BangDB}, and \textit{SurrealDB} exemplify this design. They offer first-class PG and document support and 
Native TS support is provided by \textit{BangDB} and \textit{SurrealDB} and is planned for \textit{ArcadeDB}, whereas \textit{ArangoDB}/\textit{OrientDB} rely on generic document collections for managing time series.  
A unified hybrid data model (D2) and temporal evolution (D3) are generally not provided,  but there is support for cross-model queries (Q1) and cross-model ACID transactions (Q3). 
%xpose cross-model queries through a single language (Q1), but a unified data model is missing (D2). However, time series are uneven:  provide native TS ingestion/analytics. \textit{ArcadeDB} lists TS as planned. 
ArangoDB, OrientDB, and ArcadeDB provide graph analytics (e.g., PageRank, SSSP) and aggregation capabilities for large graphs.
%and aggregations. Time-series on \textit{ArangoDB}/\textit{OrientDB} rely on generic collections (D1 for TS is absent). 
%Native temporal semantics for PG (D3) and dedicated T-indices remain rare, and retention policies are more limited than those of the other architecture categories. However, MM leads in ACID cross-model (Q3) support (ArangoDB), single-statement cross-model queries, and operational simplicity. 
An example cross-model query with ArcadeDB is provided in \cite{rep_queries}.

\section{Open challenges}
\label{sec:challenges}

 %This pattern explains why, in the heatmaps, polyglot and NN-MM sometimes “win” on coverage, whereas N-MM emphasizes transactional coherence and uniform operations over breadth of TS tooling.
\begin{figure}[t]
    \centering
    \includegraphics[width=1\linewidth]{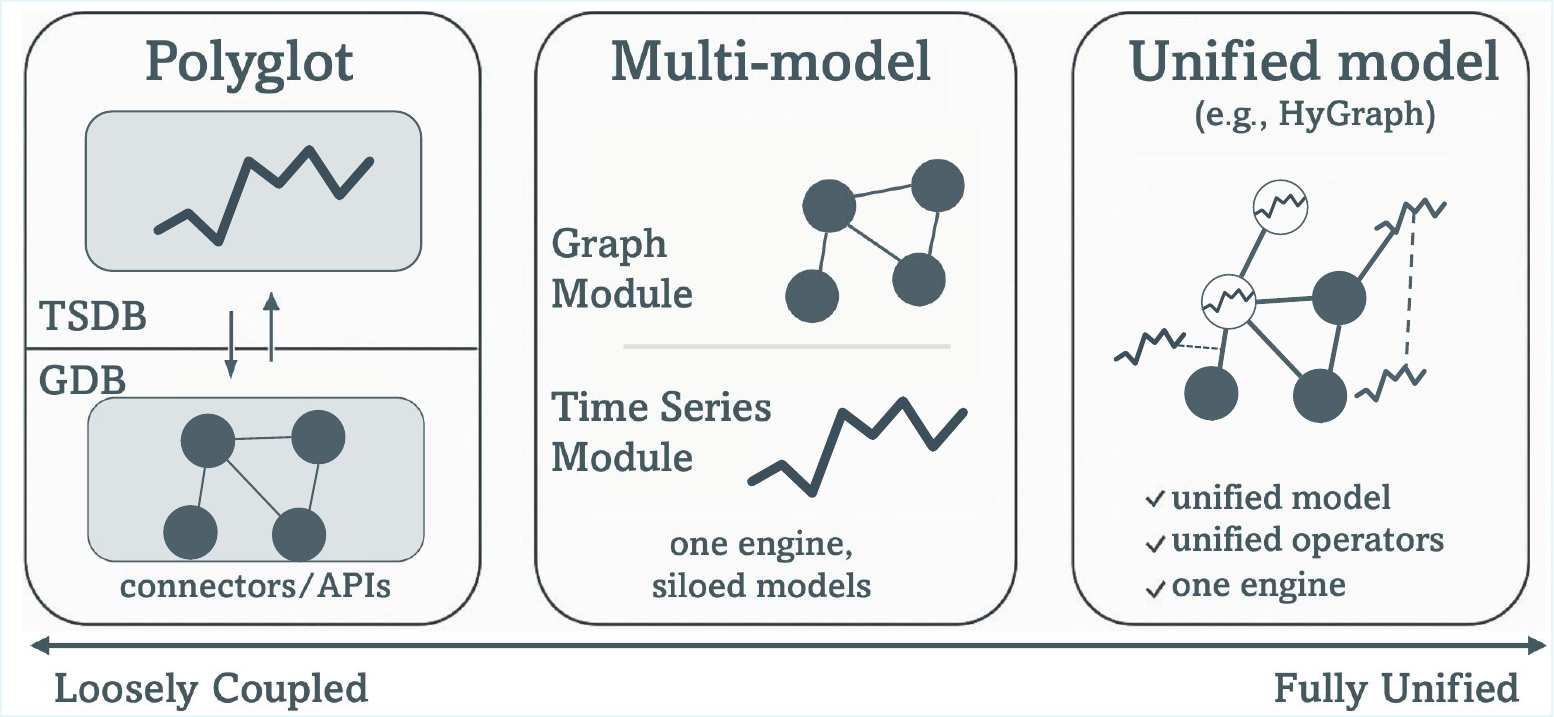}
    \caption{Integration levels for Graphs and Time series}
    \label{fig:vision}
   
\end{figure}

The heatmap in \Cref{fig:heatmap} summarizes the extent to which the surveyed systems align with our modeling, querying, and implementation criteria. Overall, no system satisfies all requirements. The best overall coverage is achieved by the E-SM system  Oracle (approximately 73\% of the defined criteria), followed by the multi-model system BangDB (about 67\%) and PostgreSQL with its extensions (62\%). The polyglot research prototype TimeTravelDB also has strong support for the criteria (about 70\%), but, of course, lacks the maturity of the former systems. Oracle offers commercial solutions; BangDB provides a community edition; however, PostgreSQL is available free of charge.

Despite the variety of existing systems with support of  graph and time series data, 
%Some requirements were not included in the tables because no system currently supports them. Thus, we identify the \textbf{following limitations}:
there are still \textbf{significant gaps} in the posed requirements and implementation features so that a 
a true unification of the two data models has not yet been achieved. 
%Some requirements were not included in the tables because no system currently supports them. Thus, we identify the \textbf{following limitations}:
Except for the support of time series properties in graphs in two research prototypes there is no support of 
 a \textbf{unified hybrid data model} (requirement  D2)  that can combine graph and time-series data in one composite data model. Therefore, a truly hybrid query language is still missing, with full support for \textbf{bidirectional data model transformations} (Q2), import/export of hybrid data, and implementation features such as composite graph/time-series indexing (P2).   
 %that allows to transform a single model graph/time-series to the composite model for hybrid analysis, to transform the composite model back to the single model are largely missing. 
 For example, consider a query that must return graph paths subject to specific time-series constraints, such as temporary spike patterns on either entities or links. With existing solutions, one must first manually link time series to their entities, then retrieve all possible paths via graph querying, and finally check, for each path, whether the condition holds in the time-series data. Such a solution would require excessive time and computation, whereas in a unified environment, one can simultaneously check time-series conditions and filter out invalid paths directly.

%\textbf{ Hybrid data import and export (from \emph{Q3})}
%Almost none of the surveyed engines can ingest a dataset that already combines graphs and time series and load it into a hybrid model in one step with consistent semantics, as well as exporting graphs and time series extracted from the composite model for compatible exportation with existing systems matters, are covered by only 2\% of the total of surveyed systems. The only existing solution is offered by TimeTravelDB, which accepts JSON files; however, processing is time-consuming (more than two hours to load our dataset~\cite{nyc_bike_data_2024} with 2,213 nodes, 5,626 edges, and 8,034,412 time-series. \textbf{Composite indexing mechanisms (\emph{P2})} that enable indexing of graph elements with temporal properties remain a standard limitation across industrial and academic solutions. In contrast, several systems provide temporal or time-series indexes for single-model workloads, composite indexing mechanisms that jointly index graph structure and temporal dimensions (e.g., for hybrid pattern matching or time-constrained traversals) are almost absent. Existing industrial and academic systems, therefore, still struggle with deep cross-model operators and cost-based optimization across heterogeneous runtimes~\cite{DBLP:journals/pvldb/KiehnSGPWWPR22,lu2019multi}.

As illustrated in \Cref{fig:vision}, the different integration architectures and their implementing systems vary in the extent to which they unify graph and time-series data. At one end,   
polyglot persistence systems loosely integrate systems of two (or more)  data models but need a complex middleware to combine the data. Multi-model systems and E-SM systems achieve a closer integration within a single DBMS and its extensions (e.g., a graph database management system and a time-series database management system) via external connectors or APIs.
but still treat each model separately with no unified data model and limited cross-model expressiveness.
%and lack a clear standard definition of the multi-model concept.
%Still, they lack shared semantics, making querying and reasoning across models cumbersome. In the middle, multi-model systems provide tighter integration by supporting multiple data models within a single engine and runtime,  However, a polyglot system with an explicit, unified hybrid model (such as TimeTravelDB) demonstrates that deeper integration is possible, but this has so far been realized only in research prototypes.
At the far end lies the vision of a truly unified data model, in which both graph and time-series data are treated as first-class citizens within a unified representation with shared semantics, indexing, and querying capabilities. The HyGraph effort~\cite{ammar2025towards} aims at providing such a solution.

%Taken together, the heatmap shows that current systems either excel along one axis (either strong graph capabilities or strong time series capabilities) or provide partial bridges between models, but none offer the level of deep, semantically unified integration required by truly hybrid workloads.
%This is precisely the gap that HyGraph aims to fill. By proposing a unified and expressive data model that integrates both structural relationships graphs) and temporal dynamics (time series), HyGraph offers additional capabilities to address the complexities of hybrid workloads and to query hybrid data. Our work underscores the need for this level of integration and demonstrates that current systems fail to adequately address the complexities of hybrid workloads. 

 %Among surveyed systems, none attains full T3. Our vision paper~\cite{ammar2025towards} motivates this target and outlines the required model, operators, and timeliness guarantees; current systems typically reach T1 or T2 with partial coverage of \textbf{D3}, \textbf{Q2}/\textbf{Q4}, and \textbf{SC2}.
\section{Conclusions}
\label{sec:conclusion}

%Applications  increasingly demand the combined  analysis over both graph data and time series ~\cite{shao2022pre,liu2023largest,mousavi2019stanford},  notably in machine learning ( \cite{chen2024graph,bloemheuvel2023graph}). 
%questions arise about how practical current hybrid approaches really are, given the lack of established foundations. This motivates a systematic account of how graphs and time series are coupled, and this combination should be modeled and stored.
We surveyed 20 systems across four architectural alternatives for combining graph and time-series data and comparatively evaluated them against a set of key requirements and implementation characteristics.
The proposed methodology can be applied to further systems that could not be included here. The quantitative evaluation can help readers to understand the current state of art and to find the most promising systems for a particular use case or research direction. 

While there are no complete solutions that cover all requirements, we found that some commercial DBMS already provide reasonable support, e.g., for evaluating both types of data in a single query. This holds especially for E-SM systems such as Oracle and PostgreSQL, as well as for multi-model approaches such as BangDB. What is currently lacking is support for a unified hybrid data model that enables closer integration of both data types, with the backing for hybrid operators and data transformations. 
%Our overview provided help readers evaluate current options and understand the central trade-offs: integration depth vs the maturity and robustness of each data model: deeper multi-model systems designs unify types, hybrid operators, and cross-model consistency; while shallower single model and extended single-model systems retain more advanced PG/TS features, ecosystems, but weaken hybrid capabilities. Polyglot systems fall between, depending on how deeply the two data models are combined, but that can also add complexity.
In future work, we plan to design a benchmark for hybrid graph and time-series processing that enables quantitative evaluation of selected systems. 
%evaluates the full spectrum of graph time-series integration methods, helping to close the remaining gap to the unified model.

%Systems cluster along two axes: where time lives in the model (bitemporal/event/transaction) and how time series is coupled to the graph (native vs federated)

\balance

\bibliographystyle{abbrv}

\bibliography{bibliography}
\end{document}